\documentclass[onecolumn]{IEEEtran}

\usepackage{epsfig}
\usepackage{color}
\usepackage[normalem]{ulem}
\usepackage[usenames,dvipsnames]{xcolor}
\usepackage{amsmath, amsthm, amssymb}
\usepackage{url}
\usepackage{graphicx}
\usepackage{multicol}
\usepackage{setspace}
\newcommand{\reqn}[1] {(\ref{#1})}

\newcommand{\be}{\begin{eqnarray}}
\newcommand{\ee}{\end{eqnarray}}
\newcommand{\beqs}{\begin{eqnarray*}}
\newcommand{\eeqs}{\end{eqnarray*}}

\newtheorem{thm}{Theorem}
\newtheorem{lemma}{Lemma}

\title{Performance Bounds\\ for Grouped Incoherent Measurements\\ in Compressive Sensing}

\author{Adam C. Polak, Marco F. Duarte,~\IEEEmembership{Senior Member,~IEEE,} and Dennis L. Goeckel,~\IEEEmembership{Fellow,~IEEE,}
\thanks{Portions of this work have previously appeared at the IEEE Statistical Signal Processing Workshop (SSP)~\cite{Polak12} and in an accompanying technical report~\cite{PolakTR}.
This paper is based in part upon work supported by the National Science
Foundation under grants CNS-0905349 and ECCS-1201835.}%
\thanks{The authors are with the Department of Electrical and Computer Engineering, University of Massachusetts, Amherst, MA 01003. E-mail: {\tt \{polak,goeckel,mduarte\}@ecs.umass.edu}}}

\begin{document}
%
\maketitle

\begin{abstract}

Compressive sensing (CS) allows for acquisition of sparse signals at sampling rates significantly lower than the Nyquist rate required for bandlimited signals. Recovery guarantees for CS are generally derived based on the assumption that measurement projections are selected independently at random. However, for many practical signal acquisition applications, including medical imaging and remote sensing, this assumption is violated as the projections must be taken in groups. In this paper, we consider such applications and derive requirements on the number of measurements needed for successful recovery of signals when groups of dependent projections are taken at random. We find a penalty factor on the number of required measurements with respect to the standard CS scheme that employs conventional independent measurement selection and evaluate the accuracy of the predicted penalty through simulations.

\end{abstract}

\section{Introduction}\label{introduction}
Modern receivers and sensors need to process an enormous amount of bandwidth to satisfy  continuously growing demands on communication and sensing systems. Due to the complexity and high power consumption of hardware at large bandwidths, a number of innovative approaches for signal acquisition have
recently emerged, including a class based on compressive sensing (CS). In CS approaches, the full signal bandwidth
is not converted, hence avoiding the costly hardware; rather, prior knowledge of a concise signal model allows the recovery to focus only on signal aspects relevant to feature extraction.
In particular, if there exists a basis in which a signal of interest can be represented sparsely (i.e., it can be fully characterized with a small number of coefficients), then it is possible to obtain all information needed for successful reconstruction of the signal from a relatively small number of randomized incoherent measurements \cite{Cand06}. This number is often much smaller than the number of samples implied by the Nyquist sampling rate for representation of all bandlimited signals.

\begin{figure}[t]
\begin{center}
\includegraphics[width=2.5in] {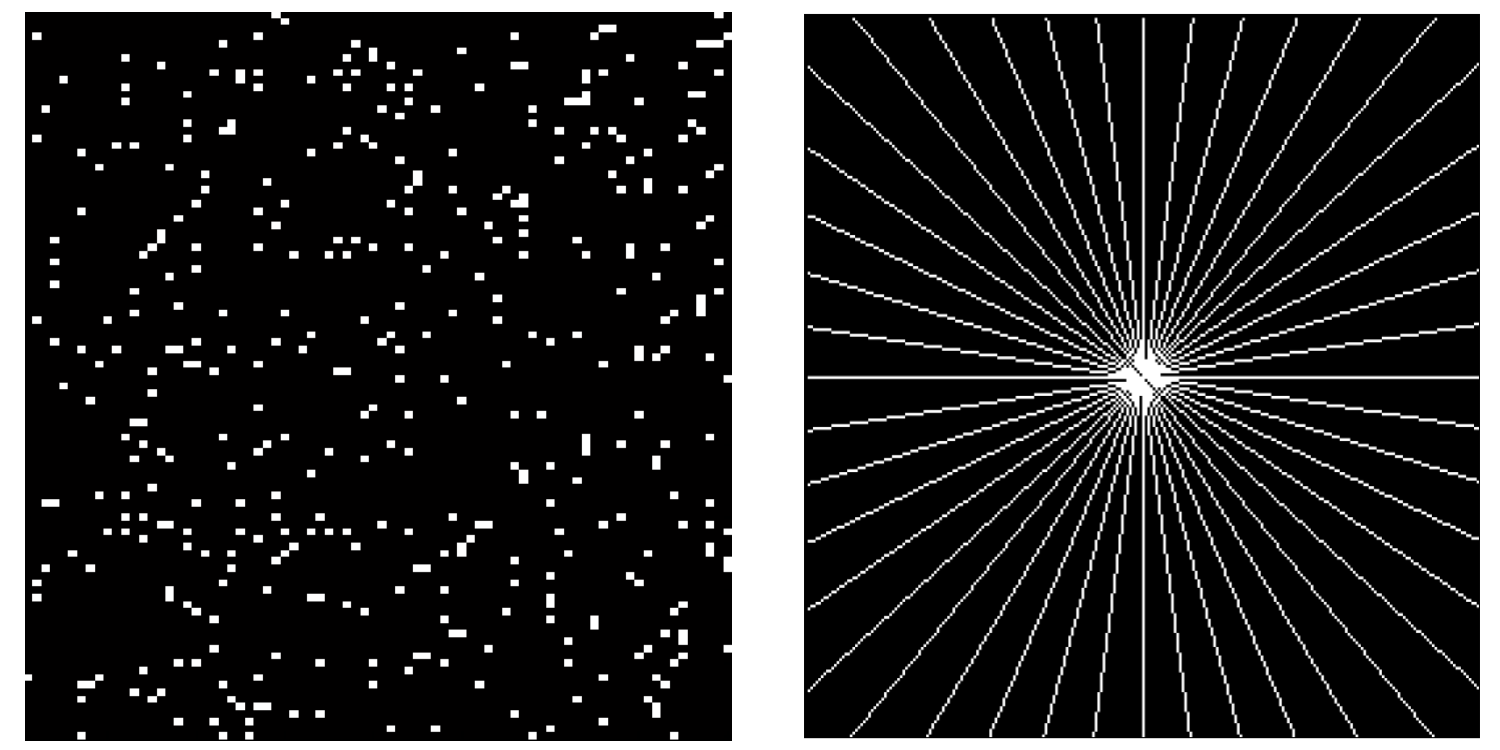}
\end{center}
\caption{\sl Left: Independently random 2-D sampling. Right: Radial acquisition trajectories used for MRI, which group measurement selections into slices of the 2-D Fourier domain. }
\label{MRI}
\end{figure}

Most CS contributions assume independent randomness in the measurement projections that is exploited to derive bounds on the number of projections needed for successful recovery. However, for many practical signal acquisition applications, this assumption is violated as the projection measurements must be selected in groups. As an example, consider Magnetic Resonance Imaging (MRI), where the measurements in the 2-D Fourier space cannot be taken at random but need to follow sampling trajectories that satisfy hardware and physiological constraints; for example, the radial acquisition trajectories of MRI shown in Figure~\ref{MRI} are known to be especially suitable for high-contrast objects. Using such sampling trajectories clearly introduces structure into the measurement process and hence violates a key assumption underlying the standard analysis of CS schemes.

In this work, we derive bounds on the number of measurements needed for successful recovery of signals when the random projection measurements are structured into predefined groups.
We introduce a metric that upper bounds the multiplicative penalty on the number of required measurements introduced by grouping with respect to conventional CS acquisition employing independently random measurement selection. The metric is dependent on the sparse signal support and might be useful in the design of many practical signal acquisition systems with grouped measurement structures. While this metric cannot currently be evaluated in a closed form, we employ a computationally feasible method that provides lower and upper bounds on its value. We also evaluate via simulations the penalty predicted by the proposed metric.

The remainder of the paper is organized as follows. Section \ref{background} provides a brief overview of the incoherent measurement scheme and compressive sensing framework, and it introduces the concept of grouped incoherent measurements together with its potential applications. Section \ref{Performance} outlines the theoretical characterization of the performance of the grouped incoherent measurement scheme, which is the main contribution of this paper. Section \ref{simulations} presents numerical results that verify the utility of the theoretical performance bounds from Section \ref{Performance}. Finally, Section \ref{conclusions} concludes the paper.

\section{Background}\label{background}
\subsection{Compressive Sensing}\label{compressive_sensing}
Consider the acquisition of an $N \times 1$ signal vector $\underline{x}$.
Assume that $\underline{x}$ is known to be sparse in some basis; that is,
we say the signal $\underline{x}$ is $K$-sparse for some integer $K$ if $\underline{x}$
has a representation $\underline{c} = U^H\underline{x}$ having only
$K$ non-zero entries in some known orthonormal basis $U$, although the value and location of those non-zero entries may be
unknown.
In the CS framework, we acquire the $M \times 1$ output $\underline{y} =
\Phi \underline{x} $, for some $M \ll N$, where $\Phi$ is the measurement matrix. According to CS theory, given certain constraints on $\Phi$ and $M$, $\underline{x}$ can be reconstructed from $\underline{y}$ with high probability.

\subsection{Incoherent Measurements}
Given an orthonormal measurement basis $V$, a $K$-sparse signal $\underline{x}=U\underline{c}$, sparse in some known orthonormal basis $U$ can be reconstructed successfully from a set of $M$ independently drawn random samples $\Omega \subseteq \{1,\ldots,N\}$ of $y=V^HU \underline{c}$ with probability not lower than $1-\delta$, for any $\delta>0$, as long as the number of samples is large enough. Define $A = V^HU$ and denote by $A_{\Omega}$ the matrix built from the $M$ rows of $A$ corresponding to the index set $\Omega$. Define the coherence $\mu(A)$ of the matrix $A$ as
$\mu(A)=\max_{i,j} |A(i,j)|$, which has range $\mu(A)\in[\frac{1}{\sqrt{N}},1]$~\cite{Cand06}.
A pair of bases $V$ and $U$ for which the minimal value of $\mu(A)$ is achieved is referred to as a {\em perfectly incoherent} pair of bases.

When the elements of $\Omega$ are drawn independently at random, it can be shown that the number $M$ of measurements required for successful recovery of  sparse $\underline{x}$ depends on the coherence of the matrix $A$.
\begin{thm}\label{theorem_Candes} \cite{Cand06} Let A be an $N\times N$ orthogonal matrix ($A^HA = I$) with coherence $\mu(A)$. Fix an arbitrary subset $T$ of the signal domain. Choose a subset $\Omega$ of the measurement domain of size $|\Omega|=M$ and a sign
sequence $z$ on $T$, both uniformly at random over all possible choices. Suppose that
\begin{equation}\label{Candes_requirement_1}
M\geq \textrm{Const} \cdot N \mu^2(A) |T| \log(N/\delta).
\end{equation}
Then with probability exceeding $1-\delta$, every signal $\underline{c}_0$ supported on $T$ with signs matching $z$ can be recovered from $y=A_{\Omega}\underline{c}_0$ by solving the linear program
\begin{equation}
\min_{\underline{c}} ||\underline{c}||_1 \quad\quad\quad s.t. \quad\quad  A_{\Omega}{\underline{c}}= A_{\Omega}\underline{c}_0.
\label{l1}
\end{equation}
\end{thm}
\noindent Theorem~\ref{theorem_Candes} shows that the number of measurements required for successful recovery of a sparse signal scales linearly with the signal's sparsity, but only logarithmically with its length, as long as $V$ and $U$ are perfectly incoherent.

\subsection{Grouped Incoherent Measurements}\label{introduce_G_i}
In certain applications, the assumptions of Theorem~\ref{theorem_Candes} are violated as measurements must be taken in groups instead of independently at random.
More specifically, divide the set of $N$ rows of $A$ into $N/g$ disjoint groups $G_i$, $i=1,\ldots,N/g$, of size $g$ each. Note that it will still be possible to take a set of measurements $\Omega$ for a signal, following Theorem~\ref{theorem_Candes}, by selecting $M/g$ groups out of the $N/g$ groups available, independently at random.\footnote{We assume that $g$ divides both $M$ and $N$ for simplicity.} We say that such a process provides a {\em grouped incoherent measurement scheme}. Grouped incoherent measurement schemes can be seen as a generalization of the standard incoherent measurement scheme used in Theorem~\ref{theorem_Candes} by setting $g=1$.

\subsection{Example Applications}

{\em Interference-Robust Compressive Wideband Receiver}: One important example application for a grouped incoherent measurement scheme is an interference-robust compressive wideband receiver. If a large communication bandwidth is employed, interference is nearly always present. More importantly, it is common for the signal of interest to be buried in an interferer that is orders of magnitude stronger. This might force the receiver's RF front end into the nonlinear range and cause intermodulation distortion that makes the interference cancellation methods based on interference null space projection~\cite{Dave2009} ineffective. As an alternative, we may opt to perform sampling only at times in which the RF front end is not saturated and exhibits linear behavior, e.g., at times when the interferer's value is small \cite{Jackson13, Polak13}. A typical interferer is modulated; therefore, while its first few zero-crossings can be considered as random, the remaining set of subsequent zero-crossings are dictated by the frequency of the interferer's carrier. Therefore, a sampling approach that aims to operate within the linear region of the RF front end results in a grouped incoherent measurement scheme, in effect providing an interference-robust compressive wideband receiver.

{\em Medical Imaging}: There are a multitude of medical imaging applications that rely on tomography principles, where CS can be applied to reduce the number of measurements required for accurate image recovery~\cite{Lust2008}; common examples include MRI and computed axial tomography (CAT). In tomographic imaging, the 2-D image measurements obtained via a tomographic scan correspond to samples of the Radon transform of the image. These samples can be grouped by orientation and processed in groups via the discrete Fourier transform. According to the projection slice theorem, the output of this transformation provides samples of the image's 2-D discrete Fourier transform along a line running through the origin (cf. Figure~\ref{MRI}). Thus, the measurements obtained correspond to a grouped measurement in the 2-D Fourier transform domain of the image, and groups can be selected independently by selecting tomographic scan orientations independently.

{\em Multi-dimensional signals and signal ensembles}: For signals spanning many physical dimensions, such as space, time, spectrum, etc., it is often difficult to design CS acquisition devices that can calculate random projections involving all signal samples. Instead, it is commonly easier to modify the CS acquisition process so that it is applied separately to each piece of a partition of the multidimensional signal. Examples include hyperspectral imaging and video acquisition, sensor networks, and synthetic aperture radar~\cite{duar2011,Patel2010}. Consider in particular the case where the choice of measurements used for each partition comes from a single orthonormal basis and is shared among partitions, introducing structure in the measurements. For example, a compressive video camera may use the same incoherent projections on each frame in the video sequence. The resulting global measurement basis is downsampled in a group-structured fashion. The grouped incoherent measurement framework can be applied when a single orthonormal basis is used for compression of the entire multidimensional signal~\cite{duar2011}.

\section{Performance Analysis for Grouped Incoherent Measurements}\label{Performance}
\subsection{Performance Metric}\label{introduce_gamma}
\begin{figure}[t]
\begin{center}
\includegraphics[width=3.5in] {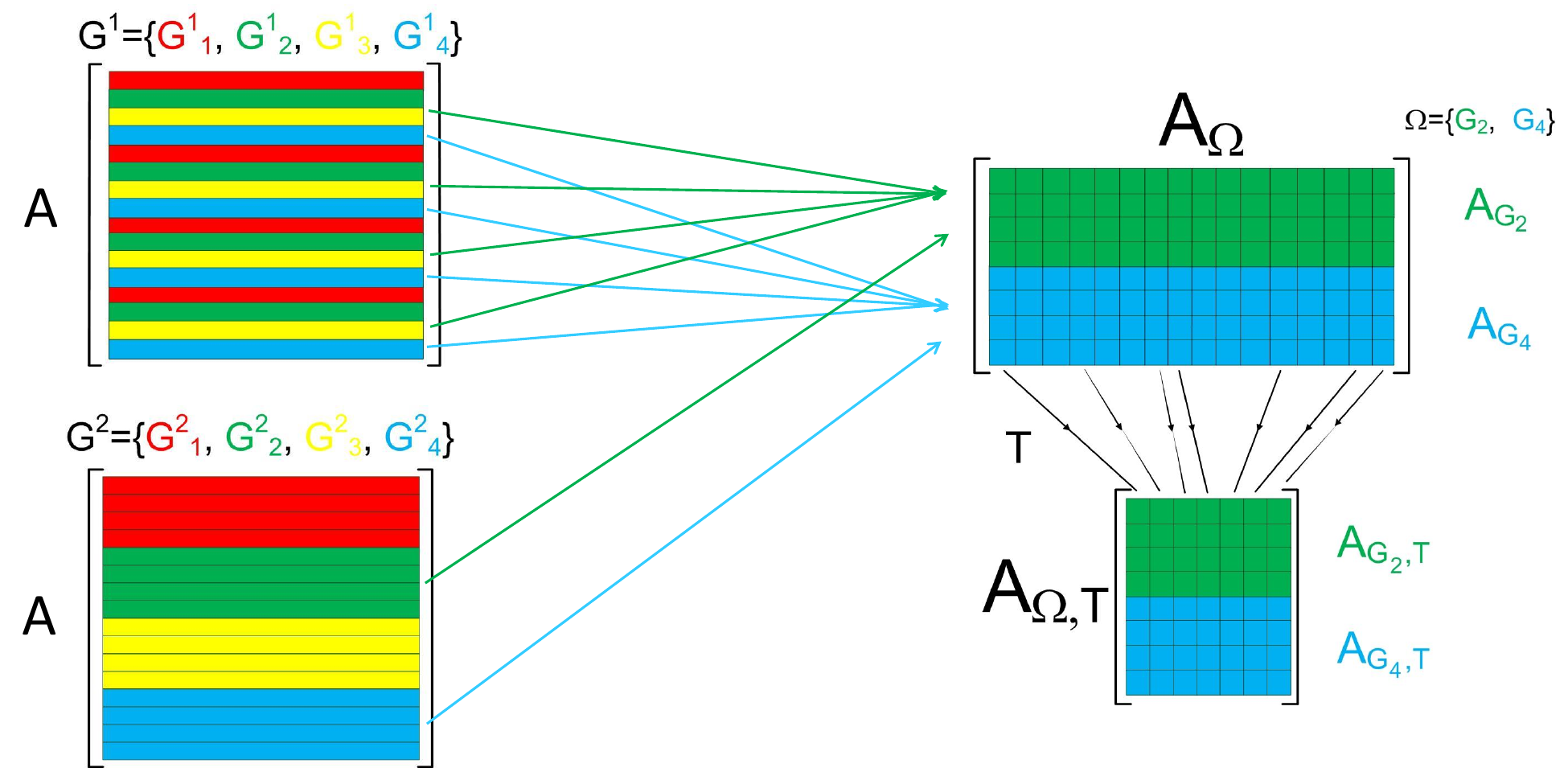}
\end{center}
\caption {Visualization of the structure of the $A_{\Omega, T}$ matrix, for $N=16$ and $g=4$, obtained by drawing two out of four groups, for two example grouping structures $G^1$ and $G^2$. In the case of $G^1$, groups are build out of samples that are separated by $N/g$ and spread over the entire sample space, whereas the in the case of $G^2$, groups are built out of adjacent samples.}
\label{grouping_strategies}
\end{figure}

The grouped incoherent measurement scheme introduced in Section~\ref{introduce_G_i} violates the assumptions of Theorem~\ref{theorem_Candes} and causes an increase of the  number of measurements needed for successful recovery of sparse signals. Such a penalty factor depends on the structure of the groups $G = \{G_1,\ldots,G_{N/g}\}$, on the product of the measurement and transformation basis $A=V^HU$, and on the set $T$ defining the sparse signal support. We define a penalty factor
\begin{equation}\label{gamma}
\gamma(A,T,G)= \max_{i\in {1,\ldots,N/g}} \left\|\overline{A_{G_iT}}\right\|_{2\to 1},
\end{equation}
where $\|M\|_{p\to q} = \max_{f} \|Mf\|_q/\|f\|_p$
denotes the $p\to q$ operator norm of the matrix $M$, $\overline{M}$ denotes the matrix $M$ after row normalization, and $A_{G_iT}$ is the submatrix of $A$ that preserves the $g$ rows corresponding to the group $G_i$ and the $|T|$ columns corresponding to the sparsity set $T$. Given the set $T$ defining the sparse support, the penalty factor $\gamma(A,T,G)$ is a measure of similarity among the rows of $A_{G_iT}$ for each $i$. For example, if the rows of $A_{G_iT}$ are equal for some $i$, we will have $\gamma(A,T,G) = g$; in contrast, if all rows of $A_{G_iT}$ are mutually orthogonal for each $i$, then we will have $\gamma(A,T,G) = \sqrt{g}$. Figure~\ref{grouping_strategies} shows the structure of the matrix $A_{\Omega,T}$ obtained by drawing two out of four groups for two example grouping structures $G^1$ and $G^2$; here $N=16$ and $g=4$.  In the case of $G^1$, groups are build out of samples that are separated by $N/g$ and spread over the entire sample space, whereas in the case of $G^2$, groups are built out of adjacent samples.

\subsection{Recovery Guarantees}\label{theory}
We now provide requirements on the number of measurements needed for successful recovery of the sparse signal $\underline{x}$ when the subset $\Omega$ of the measurement domain is built in a structured way.
\begin{thm}\label{theorem_1} Let $A$ be an $N\times N$ orthogonal matrix ($A^HA = I$) with coherence $\mu(A)$. Fix an arbitrary subset $T$ of the signal domain.
Choose a subset $\Omega$ of the measurement domain of size $|\Omega|=M$ as the union of $M/g$ groups from $G = \{G_1,\ldots,G_{M/g}\}$ and a sign sequence $z$ on $T$, both uniformly at random over all possible choices.
Suppose that
\begin{equation}\label{our_requirement_1}
M\geq \gamma(A,T,G)\cdot \textrm{Const} \cdot \mu^3(A) N^{3/2}  |T| \log(N/\delta).
\end{equation}
Then with probability exceeding $1-\delta$,
every signal $\underline{c}_0$ supported on $T$ with signs matching $z$ can be recovered from $y=A_{\Omega}\underline{c}_0$ by solving the linear program (\ref{l1}), for any $\delta>0$.
\end{thm}
The theorem shows that for a perfectly incoherent measurement and sparsity bases, $\gamma(A,T,G)$ provides a multiplicative penalty on the number of measurements necessary for successful signal recovery due to the grouped structure of the incoherent measurement selection. Note that for a group size $g=1$ and for perfectly incoherent pair of bases $V$ and $U$ our result coincides with Theorem~\ref{theorem_Candes} as it is equivalent to drawing elements of $\Omega$ uniformly at random.

\begin{IEEEproof}
In the following,  we will prove the result of Theorem~\ref{theorem_1} with a small modification on the distribution of the submatrices: instead of a uniform distribution among all subsets $\Omega$ containing $M/g$ out of the $N/g$ available groups, we pose an independent Bernoulli selection for each group submatrix $G_i$, $i=1,\ldots,N/g$, belonging in $\Omega$ with selection probability $P(\delta_i=1) = M/N$. This independent model results in the expected number of selected groups being equal to $M/g$. Furthermore, one can show that since the probability of failure is a non-increasing function of the size $M$ of the set $\Omega$, the probability of failure under the uniform distribution used in Theorem~\ref{theorem_1} is upper-bounded by a constant times the probability of failure under the independent selection model used in the proof (a property dubbed {\em poissonization} in~\cite{Cand06_2}). Thus, the effect of the conversion of the subgroup selection model is a constant multiplicative factor in the required number of measurements, which is accounted for by the constants in \reqn{Candes_requirement_1} and \reqn{our_requirement_1}.

Following the argument of~\cite{Cand06}, one can show that the signal $\underline{c}_0$ is the unique solution to \reqn{l1} if and only if there exists a dual vector $\pi  \in \mathbb{R}^N$ that has following properties:
\begin{itemize}
\item $\pi$ is in the row space of $A_{\Omega}$,
\item $\pi(t)=\textrm{sign} \{\underline{c}_0(t)\}$ for $t\in T$,
\item $|\pi(t)|<1$ for $t\in T^c$.
\end{itemize}
As in \cite{Cand06}, we consider the candidate
\begin{equation}
\pi =A_{\Omega}^HA_{\Omega T}(A_{\Omega T}^HA_{\Omega T})^{-1}z_0,
\end{equation}
where $z_0$ is a $|T|$-dimensional vector whose entries are the signs of $\underline{c}_0$ on $T$. To prove Theorem \ref{theorem_1} we need to show that under its hypothesis:
($i$) $A_{\Omega T}^HA_{\Omega T}$ is invertible and ($ii$) $|\pi(t)|<1$ for $t\in T^c$.
We begin by showing that $A_{\Omega T}^HA_{\Omega T}$ is invertible with high probability given the requirement \reqn{our_requirement_1} on $M$. The following theorem is proven in Appendix~\ref{first_proof} and shows that if $M$ is large
enough then, on average, the matrix $A_{\Omega T}^HA_{\Omega T}$ does not deviate much from $\frac{M}{N}I$, where $I$ is the identity matrix.

\begin{thm}\label{theorem1_2}
Fix an arbitrary subset $T$ of the signal domain. Define $N/g$ index groups $G = \{G_1,\ldots,G_{N/g}\}$ of the measurement domain, each of size $g$, and draw each group independently at random with probability $M/N$ into a set $\Omega$. If
\begin{equation}
M\geq  \frac{28}{3} \cdot \gamma(A,T,G) \cdot N  \cdot \mu^2(A)\cdot |T| \log \left(\frac{|T|}{\delta} \right),
\end{equation}
with $\gamma(A,T,G)$ introduced in \reqn{gamma}, then
\begin{equation}
P\left(\left\|\frac{N}{M}A_{\Omega T}^HA_{\Omega T}-I\right\|\geq \frac{1}{2}\right)<\delta,
\end{equation}
where $\left\| \cdot \right\|$ denotes the spectral norm
\begin{equation}\label{spec_norm_def}
\left\| Y \right\|= \sup_{\|f_1\|_2 = \|f_2\|_2 = 1}|\langle f_1,Y f_2\rangle |.
\end{equation}
\end{thm}

Theorem~\ref{theorem1_2} shows that if $M$ is large enough, then $A_{\Omega T}^HA_{\Omega T}$ is invertible with high probability. We continue by proving that $|\pi(t)|<1$ for $t\in T^c$. Following the techniques in \cite{Cand06}, we use the following three lemmas, proven in the appendix.

\begin{lemma}\label{lemma1}
Denote by $v^0$ a row of the matrix $A_\Omega^H A_{\Omega T}$ indexed by $t_0\in T^C$. Then
\begin{equation}
E||v^0||^2<\frac{M}{\sqrt{N}} \mu^3(A) |T| \gamma.
\end{equation}
\end{lemma}

\begin{lemma}\label{lemma2}
Define
\begin{equation}\label{define_sigma}
\bar{\sigma}^2:= \gamma \cdot \mu^{2}(A) \frac{M}{N} \cdot \max\left\{1, \sqrt{\gamma} \mu^{3/2}(A) N^{3/4} |T|/ \sqrt{M}\right\}.
\end{equation}
For \hspace{2.1in} $ 0<a\leq \frac{\sqrt{M}}{ {\mu(A)} \sqrt{N \gamma |T|}}$ if $\frac{\sqrt{\gamma}\mu^{3/2} N^{3/4}(A)|T|}{\sqrt{M}}<1$ \\
and \hspace{1.8in} $ 0<a\leq \left(\frac{{M}}{{\gamma} \mu(A) \sqrt{N}}\right)^ {1/4}$ if $\frac{\sqrt{\gamma}\mu^{3/2} N^{3/4}(A)|T|}{\sqrt{M}}\ge 1$, \\
we have
\begin{equation}
P\left(||v^0||>\mu^{3/2}(A)N^{-1/4}\sqrt{\gamma M |T|}+a\bar{\sigma}\right)<3 e^{-\kappa a^2}
\end{equation}
for some positive constant $\kappa$.
\end{lemma}

\begin{lemma}\label{lemma3}
Let $w^0=(A_{\Omega T}^HA_{\Omega T})^{-1}v^0$. With the notations and assumptions of Lemma \ref{lemma2} we have:
\begin{equation}
\begin{split}
P\left(\sup_{t_0\in T^c}||w^0|| \geq 2  N^{3/4} \mu^{3/2}  \sqrt{\frac{\gamma |T|}{M}}+ \frac{2 N a\bar{\sigma}}{M} \right) \leq 3 e^{-\kappa a^2} + P\left( ||A_{\Omega T}^HA_{\Omega T}|| \leq \frac{M}{2N} \right).
\end{split}\label{lemma3_inequality}
\end{equation}
\end{lemma}

\noindent Finally we will use \cite[Lemma 3.4]{Cand06}, reproduced below.
\begin{lemma}\label{lemma4} Assume that $z(t)$, $t \in T$ is an $i.i.d.$ sequence of symmetric Bernoulli random variables.
For each $\lambda> 0$, we have
\begin{equation}\label{lemma4_inequality}
P\left(\sup_{t\in T^c} |\pi(t)|>1 \right)\leq 2Ne^{-1/2\lambda^2}+P\left( \sup_{t\in T^c} ||w^0||>\lambda \right).
\end{equation}

\end{lemma}

Now that all lemmas are in place, we are ready to prove Theorem \ref{theorem_1}. If we pick $\lambda=2 N^{3/4} \mu^{3/2} \sqrt{\gamma|T|/M}+2N a\bar{\sigma}/M$ in \reqn{lemma4_inequality}, from \reqn{lemma3_inequality} and \reqn{lemma4_inequality} we get
\begin{eqnarray}
P  \left(\sup_{t\in T^c} |\pi(t)|>1 \right)& \leq & 2Ne^{-1/2\lambda^2}+Ne^{-\kappa a^2} \label{gg} + P\left( ||A_{\Omega T}^HA_{\Omega T}||\leq M/2N) \right).
\end{eqnarray}
For the right hand side of \reqn{gg} to be smaller than $ 3 \delta$ we need all three summands to be smaller than $\delta$. We now derive conditions on $\delta$ that provide this guarantee. We start with the second summand: for it to be no bigger than $\delta$ we can set $a^2$ to be
\begin{equation}\label{a_square}
a^2=\kappa^{-1}\log(N/\delta).
\end{equation}
For the first summand to be no bigger than $\delta$, we need
\begin{equation}\label{lambda_square}
\frac{1}{\lambda^2}\geq 2\log(2N/\delta).
\end{equation}
If $\frac{\sqrt{\gamma}\mu ^{3/2}(A) N^{3/4} |T|}{\sqrt{M}}>1$, Lemma \ref{lemma2} requires
\begin{equation}\label{assumption2_1}
0<a\leq \left(\frac{{M}}{{\gamma} \mu (A) \sqrt{N}}\right)^ {1/4}.
\end{equation}
Then with $\bar{\sigma}^2$ from \reqn{define_sigma} we get
\begin{equation}\label{h}
N a\bar{\sigma}/M\leq \mu^{3/2} N^{3/4}\sqrt{\gamma |T|/M},
\end{equation}
and so
\begin{equation}\label{h_1}
\lambda \leq 4\mu^{3/2}(A) N^{3/4} \sqrt{\gamma |T|/M}.
\end{equation}
Reorganizing terms, we obtain
\begin{equation}
\frac{1}{\lambda^2}\geq \frac{M}{16\mu^3(A) N^{3/2} \gamma |T|}.
\label{match}
\end{equation}
From \reqn{a_square} and \reqn{assumption2_1} we get the following bound on $M$:
\begin{equation}\label{req_0}
M\geq \gamma \mu(A) \sqrt{N} \kappa^{-2} \log^2(N/\delta).
\end{equation}
Suppose now that $\frac{\sqrt{\gamma}\mu ^{3/2}(A) N^{3/4} |T|}{\sqrt{M}}<1$. Then, with \reqn{define_sigma}, $\bar{\sigma}^2=\gamma \mu^{2}(A) \frac{M}{N}$. If $\mu(A) \sqrt{N} |T| \geq a^2$, then
\begin{equation}
N a\bar{\sigma}/M\leq \mu^{3/2} N^{-1/4}\sqrt{\gamma |T|/M},
\end{equation}
and
\begin{equation}
\lambda \leq 4\mu^{3/2}(A) N^{3/4} \sqrt{\gamma |T|/M},
\end{equation}
and thus
\begin{equation}\label{min_1}
\frac{1}{\lambda^2}\geq \frac{M}{16\mu^3(A) N^{3/2} \gamma |T|},
\end{equation}
which matches the previous condition \reqn{match}. On the other hand, if $\mu(A) \sqrt{N} |T| \leq a^2$ then
\begin{equation}
N a\bar{\sigma}/M \geq \mu^{3/2} N^{-1/4}\sqrt{\gamma |T|/M},
\end{equation}
and
\begin{equation}
\lambda \leq 4Na\bar{\sigma}/M,
\end{equation}
and thus, with $\bar{\sigma}^2=\gamma \mu^2(A)\frac{M}{N}$,
\begin{equation}\label{min_2}
\frac{1}{\lambda^2}\geq \frac{M^2}{16 N^2 a^2{\bar{\sigma}}^2}= \frac{M}{16a^2\gamma \mu^{2}(A) N}.
\end{equation}
And so with \reqn{min_1} and \reqn{min_2} we can write
\begin{equation}
\frac{M}{16 \gamma \mu^2(A) N}\min \left( \frac{1}{ \mu(A) N^{1/2} |T|} , \frac{1}{ a^2 } \right) \geq 2\log(2N/\delta),
\end{equation}
\begin{equation}
M\geq 16 \gamma \mu^2(A) N\max \left( \mu(A) N^{1/2} |T|, a^2 \right)  2\log(2N/\delta),
\end{equation}
which with \reqn{a_square} gives
{\fontsize{9}{10}\selectfont
\begin{equation}\label{req_1}
M\geq \textrm{Const} \cdot \gamma \mu^2(A) N\max \left( \mu(A) N^{1/2} |T|, \log\left( \frac{N}{\delta} \right) \right)  \log\left(\frac{N}{\delta}\right).
\end{equation}}
Due to Theorem \ref{theorem1_2}, for the third summand to be smaller than $\delta$, we need
\begin{equation}\label{req_2}
M\geq  \frac{28}{3} \cdot \gamma \cdot N  \cdot \mu^2(A)\cdot |T| \log \left(\frac{|T|}{\delta} \right).
\end{equation}
Thus from \reqn{req_0}, \reqn{req_1} and \reqn{req_2} we see that the overall requirement on $M$ is:
\begin{equation}
M\geq  \textrm{Const} \cdot \gamma(A,T,G)\cdot  \mu^{3}(A) N^{3/2} |T| \log(N/\delta),
\end{equation}
which finishes the proof of the Theorem \ref{theorem_1}.
\end{IEEEproof}

\subsection{Calculation of the Performance Metric}

For a fixed sparsity set $T$, we can obtain lower and upper bounds on the value of $\gamma(A,T,G)$ by leveraging the Pietsch Factorization theorem~\cite{Tropp2009}, which is as a basic instrument in modern functional analysis \cite{pis86}. 
\begin{thm}\label{theorem_3}
Each matrix $B$ can be factored as $B = FD$ where $D$ is a nonnegative, diagonal matrix with $\textrm{trace}(D^2) = 1$ and $\|B\|_{\infty\to 2} \le \|F\|_2 \le K_p \|B\|_{\infty \to 2}$, where $K_p$ is a constant equal to $\sqrt{\frac{\pi}{2}}\approx 1.25$ for the real field and $\sqrt{\frac{4}{\pi}} \approx 1.13 $ for the complex field.
\end{thm}
Since $\|M\|_{2\rightarrow 1} = \|M^H\|_{\infty \rightarrow 2}$, thanks to the duality of the operator norms, we can find bounds on $\gamma$ by performing Pietsch factorization of the matrices $(\overline{A_{G_iT}})^H =F_i D_i$, for $i=1,\ldots, N/g$,
where $D_i$ is a nonnegative diagonal matrix with $\textrm{trace}(D_i^2)=1$. The value of $\gamma(A,T,G)$ can then be bounded by
\begin{equation}\label{pietsch}
\frac{1}{K_p} \max_{i} ||F_i||_2 \leq \gamma(A,T,G) \leq \max_{i} ||F_i||_2,
\end{equation}

The Pietsch factorization of matrix $B$ can be performed by solving a semidefinite program \cite{Tropp2009}.

\section{Simulations}\label{simulations}

In this section, we present simulation results that justify the utility of the penalty factor $\gamma$ \reqn{gamma} as an indicator of the recovery performance of different group structures for the grouped incoherent measurement scheme. First, one-dimensional Fourier sparse signals are considered. Next, we present the dependency of the recovery performance on the penalty factor for multiple different grouping structures for images.

\subsection{Fourier-Domain Sparse 1-D Signals}

\begin{figure}[t]
\begin{center}
\includegraphics[width=3.5in] {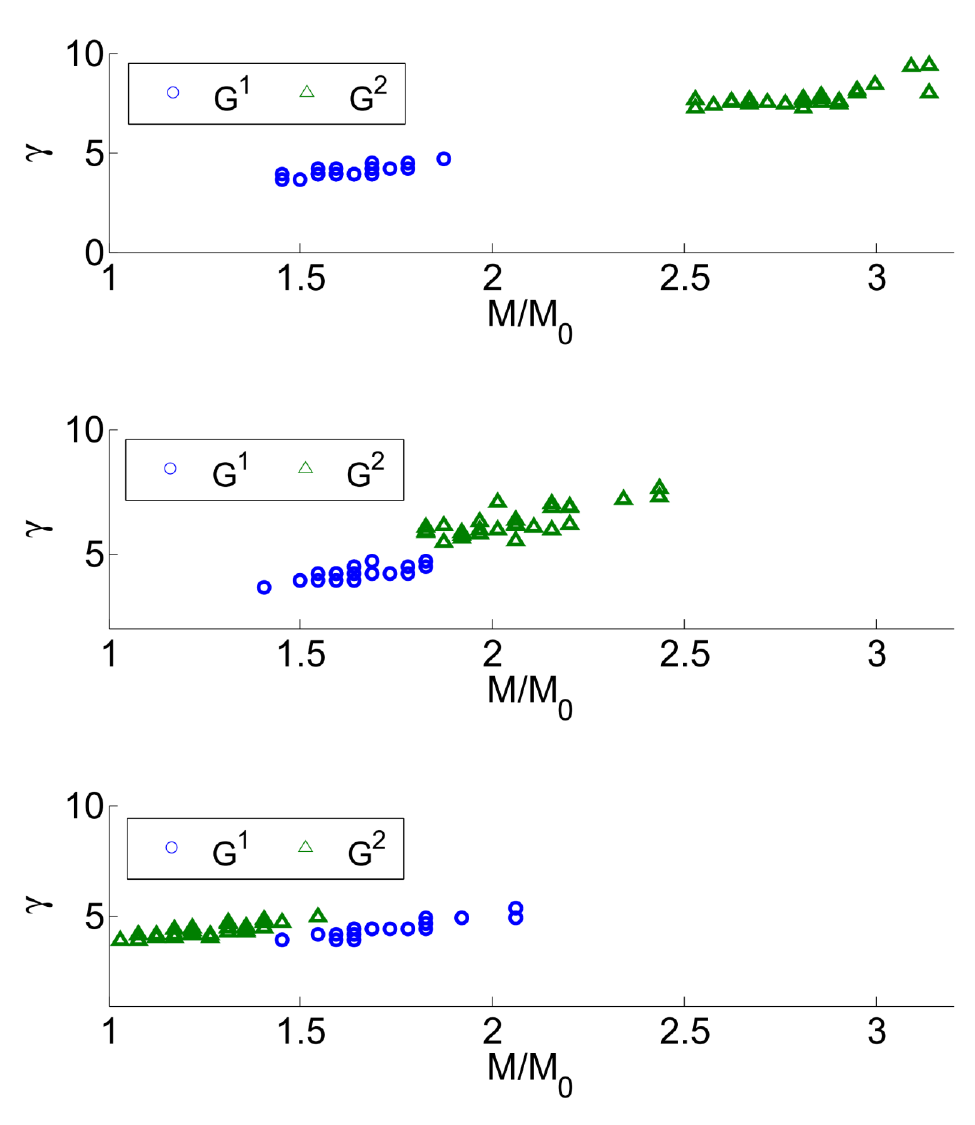}
\end{center}
\caption{ $\gamma$ vs. $M/M_0$ for group structures $G^1$ and $G^2$ for different concentrations of the nonzero Fourier coefficients of a $5\%$ sparse signal $s$. Top: a sub-band built out of two $5\%$-wide channels; middle: a sub-band built out of four $5\%$-wide channels; bottom: the entire band.}
\label{Fig1}
\end{figure}

We generate discrete signals $s$ of length $N=1100$ and sparsity $|T|=5\% \cdot N$, sparse in the frequency domain, generated as a product of an orthonormal Fourier  basis of size $N\times N$ and a sparse coefficient vector $\underline{c}$ with values of non-zero entries distributed uniformly: $\sim \mathcal{U}(-1,1)$. We evaluate two different configurations for the grouped incoherent measurements:
\begin{itemize}
\item $G^1$: 100 groups of size 11 were constructed such that the first sample of each of the groups was chosen out of the first 100 samples of $s$: $\{s[1],\ldots,s[n]\}$, and the remaining 10 samples for each group were shifted with respect to the first sample by multiples of 100. More specifically, $G^1_i=\{i,i+100,i+200,\ldots,i+1000\}$. This  configuration appears in the interference-robust compressive wideband receiver application. The first sample corresponds to a random zero-crossing of a modulated interferer. Additional samples correspond to subsequent zero-crossings of the interferer's carrier.
\item$G^2$: 100 groups of size 11 were constructed such that each group contained 11 consecutive, adjacent samples. More specifically, $G^2_i=\{s[i+(i-1)\cdot11]:s[i\cdot 11]\}$. Such configuration assumes that the samples are taken in sequential bursts.
\end{itemize}

\noindent These configurations correspond to different partitioning of the measurement domain into nonoverlapping groups, which is equivalent to partitioning of rows of the transformation matrix $A=V^HU$ into nonoverlapping groups, as visualized in the left part of Figure \ref{grouping_strategies}.

Figure~\ref{Fig1} shows the relation between the penalty factor $\gamma(A,T,G)$ from \reqn{gamma}  and the ratio between the number $M$ of samples required for successful recovery for the two described group structures and the number of samples $M_0$ required for successful recovery for random sampling. The values shown are the minimal number of measurements needed to obtain normalized recovery error $NRE = \| s- \hat{s}\| /\|s\| < 0.001$  for $99$ out of $100$ draws of the measurement groups (uniformly at random) and the values of the Fourier coefficients (from $\mathcal{U}[-1,1])$.\footnote{Throughout this section, the SPGL1 solver \cite{spgl1, spgl1_b} was used for recovery, while the CVX optimization package \cite{cvx} was used to solve a semidefinite program \cite{Tropp2009} for Pietsch factorization of the matrices $(\overline{A_{G_iT}})^H$ and subsequent calculation of the penalty factors $\gamma(A,T,G)$.} Each point of the scatter plots corresponds to a fixed signal support. We consider three different classes of signal supports: for the first two classes, the positions of the non-zero Fourier coefficients are chosen uniformly at random within a sub-band built out of two and four $5\%$-wide channels, respectively, positioned uniformly at random within the entire frequency band; we then compare their performance against the baseline of signals with unrestricted sparse supports. Figure~\ref{Fig1} shows that for the first two classes $\gamma$ was a good performance indicator; in contrast, for the last class the values of $\gamma$ misleadingly suggest that both group structures perform equally well.  This is indicative of the potential looseness of the bound provided by Theorem~\ref{theorem_1}. We believe that such looseness is characteristic of guarantees that rely on worst-case metrics, such as the coherence $\mu(A)$ and our metric $\gamma(A,T,G)$, and is compounded by the looseness in the estimate of $\gamma(A,T,G)$ obtained via Theorem~\ref{theorem_3} (of up to $1/\sqrt{\pi/2} \approx 21\%$).

\subsection{Wavelet domain sparse 2-D signals}

\begin{figure}[t]
\begin{center}
\includegraphics[width=3.3in] {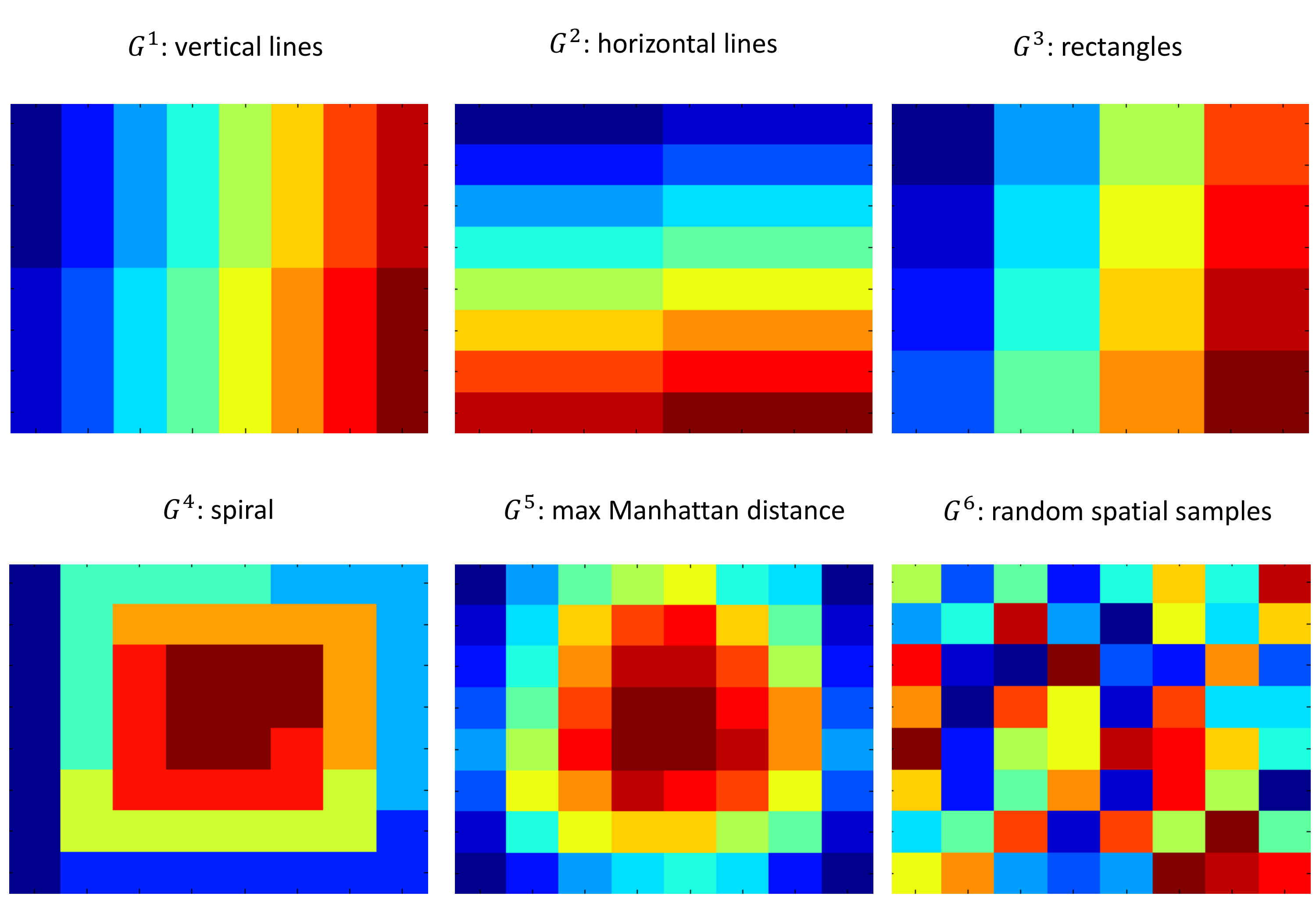}
\end{center}
\caption{Illustration of tested group structures for $8\times8$-pixel images and for a group size $g=4$, where elements of the same group are marked with the same color.}
\label{2D_groups}
\end{figure}

Next, we consider the recovery of images from grouped measurements. For different measurement trajectories (group structures), we use the penalty factor to assess the suitability of different group measurement structures to obtain successful recovery with the least number of measurements. We consider six different 2-D group structures:
\begin{itemize}
\item $G^1$: vertical lines;
\item $G^2$: horizontal lines;
\item $G^3$: $ g/2 \times 2$ rectangles;
\item $G^4$: spiral;
\item $G^5$: maximal Manhattan distance; and
\item $G^6$: groups build out of random spacial samples.
\end{itemize}
Figure~\ref{2D_groups} shows the structures for $8\times8$-pixel images and for a group size $g=4$, where elements of the same group are marked with the same color. The group structure $G^5$ was constructed as follows: the upper left pixel was chosen as the first element of the first group, and successive elements of the group were chosen from the remaining pixels to maximize the total Manhattan distance between the new element and the existing elements of the group. After all elements of the group were chosen, a new group was constructed starting with the pixel closest to the top left corner among those remaining, following the same procedure as the first group afterwards; this procedure was repeated for all other groups.

The suitability of the penalty factor $\gamma$ as an indicator of the performance of different 2-D group measurement structures was evaluated with two sets of experiments. The first experiment evaluates grouped sampling, i.e., spatial measurements. The second experiment evaluates grouped frequency-domain measurements that emulate MRI acquisition.

\subsubsection{Recovery of Satellite Terrain Images}\label{satellite}

\begin{figure}[t]
\begin{center}
\includegraphics[width=3.5in] {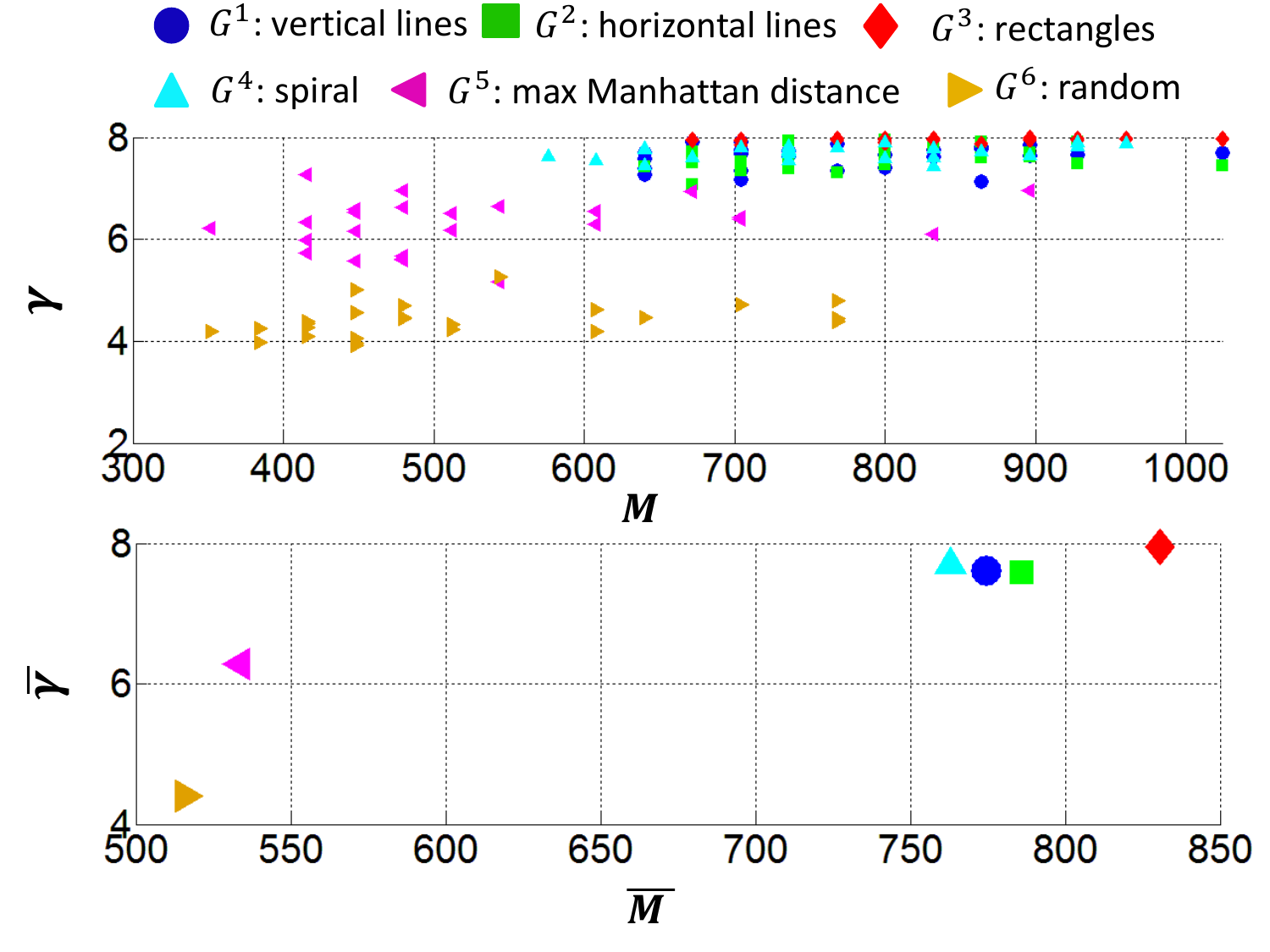}
\end{center}
\caption{Top: relationship of $M$ vs.\ $\gamma$ for the six considered group structures, for 25 low-resolution ($32\times 32$ pixels) compressed images from a satellite terrain images of areas around the town of Amherst; bottom: average value of $\gamma$ and $M$, averaged over the 25 segments.}
\label{figure_satelite}
\end{figure}

The images used in the first experiment are taken from a satellite terrain image of areas around the town of Amherst, MA that was obtained from Google Maps.  25 low-resolution ($32\times 32$ pixels) tiles are gray-scaled and compressed using wavelet transform coding to 51 coefficients. We study the recovery of these images from grouped pixel measurements under configurations $G^1$-$G^6$ with groups of size $g=8$.
Figure~\ref{figure_satelite} shows the relationship between the penalty factor $\gamma(A,T,G)$  and the number $M$ of samples required for successful recovery for each of the six group structures from Figure~\ref{2D_groups}. Each point of the top scatter plot corresponds to a single $32\times 32$-pixel tile, while each point of the bottom scatter plot shows the average values of $\gamma$ and $M$, over all of the tiles, for each of the grouped measurement configuration. In these experiments, recovery success is defined by a normalized recovery error $NRE = \| s- \hat{s}\| /\|s\| < 0.1$  for $49$ out of $50$ draws of the measurement groups, uniformly at random. The values of $M$ tested are multiples of $4\cdot g=32$.

Figure~\ref{figure_satelite} shows how the value of $\gamma(A,T,G)$ increases as a function of the number of measurements $M$ required for successful recovery until it reaches its maximal value $\gamma=g=8$ for the group structure $G^3$. The Figure~shows that the metric ${\gamma}$ can be a useful indicator of the performance for group structures of practical interest. The metric indicates a superior performance of the randomized sampling structure $G^6$, as well as the Manhattan distance-based group structure $G^5$, both of which bear out in practice. Out of the four group structures $G^1,  G^2 , G^3$ and $G^4$, characterized with continuous measurement trajectories, $G^3$ exhibited the worst performance, and the highest value of the penalty $\gamma(A,T,G)$. The recovery performance, as well as the value of $\gamma(A,T,G)$, was very similar for group structures $G^1$, $G^2$ and $G^4$. Despite similar performances for group structures $G^5$ and $G^6$  a certain level of variation of the $\gamma$ factor was observable.

\subsubsection{Recovery of MRI Images}

\begin{figure}[t]
\begin{center}
\includegraphics[width=3in] {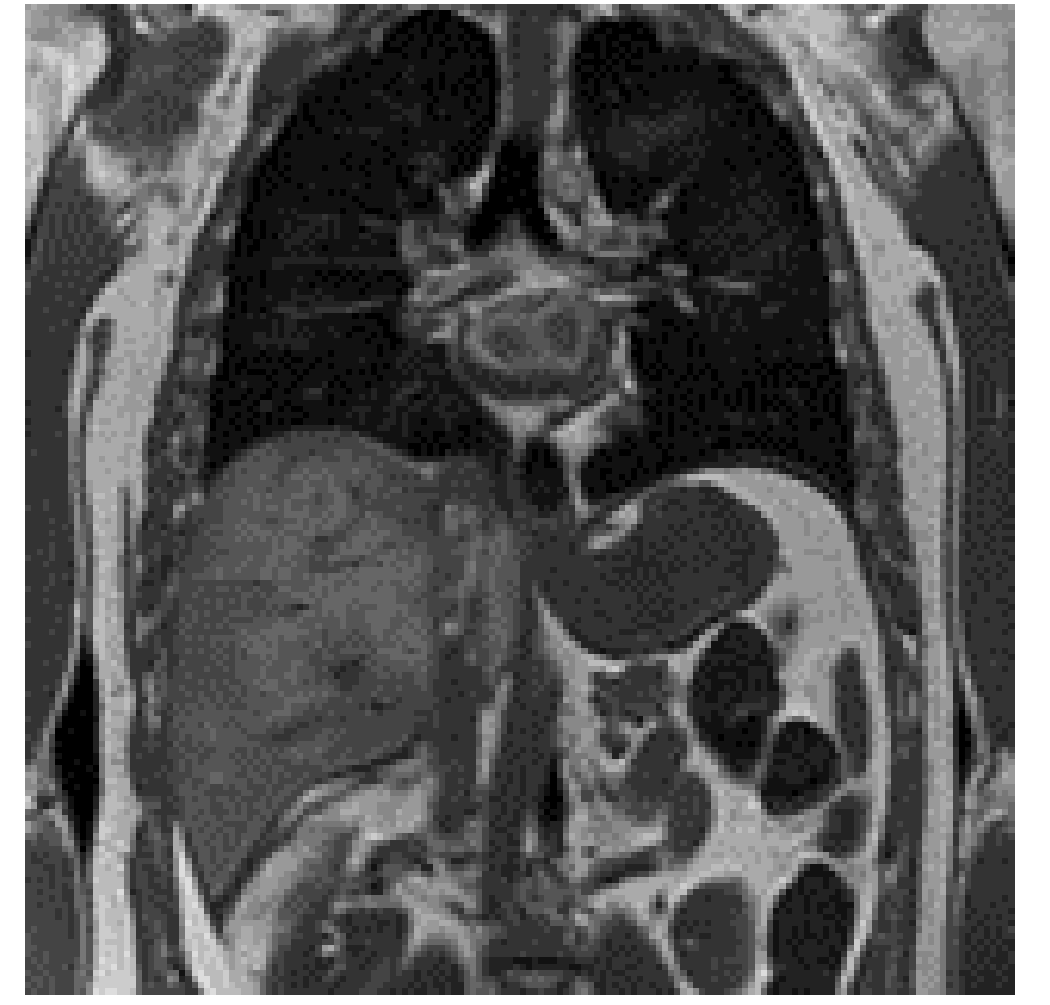}
\end{center}
\caption{$160\times 160$-pixel chest MRI image used in the experiment.}
\label{chest_MRI}
\end{figure}

\begin{figure}[t]
\begin{center}
\includegraphics[width=3.5in] {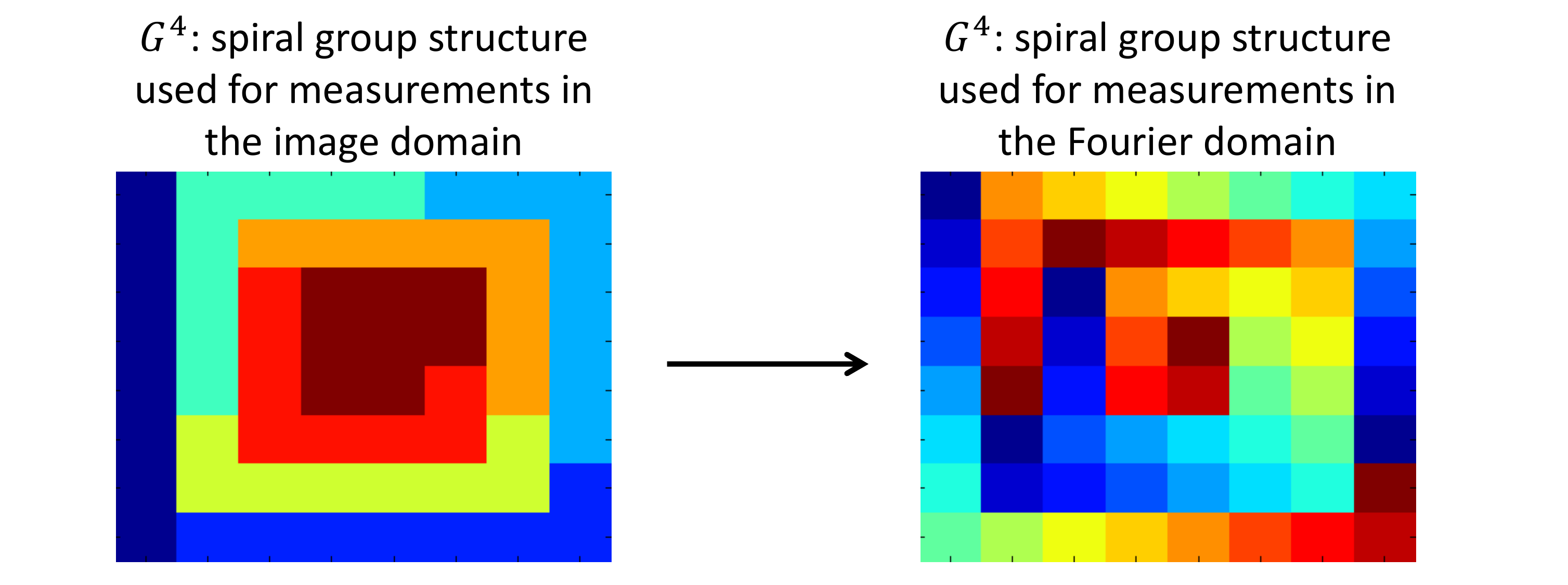}
\end{center}
\caption{ Grouped measurement structure $G^4$ used in the MRI experiments.}
\label{new_G4}
\end{figure}

In the second experiment, we study the recovery of MRI images from grouped measurements taken in the Fourier domain. 25 small-scale ($32\times32$ pixels) images were obtained as segments of an $160\times 160$ pixels chest MRI image from Figure~\ref{chest_MRI} and compressed using wavelet transform coding to 51 coefficients. The group size was again set to $g = 8$. For the MRI experiments, the spiral group structure $G^4$ shown in Figure~\ref{2D_groups}, where adjacent measurements form a spiral trajectory, was replaced with a structure where adjacent measurements in the same spiral trajectory are assigned to the different groups lexicographically and cyclically. For such a grouping structure, the measurements contributing to a given group were spread across the spectrum of the considered 2-D signal -- including both low and high-frequency measurements in each group. Figure~\ref{new_G4} visualizes the new grouping structure $G^4$ for the Fourier measurement domain of size $8\times8$ and for a group size $g=4$.

\begin{figure}[t]
\begin{center}
\includegraphics[width=3.5in] {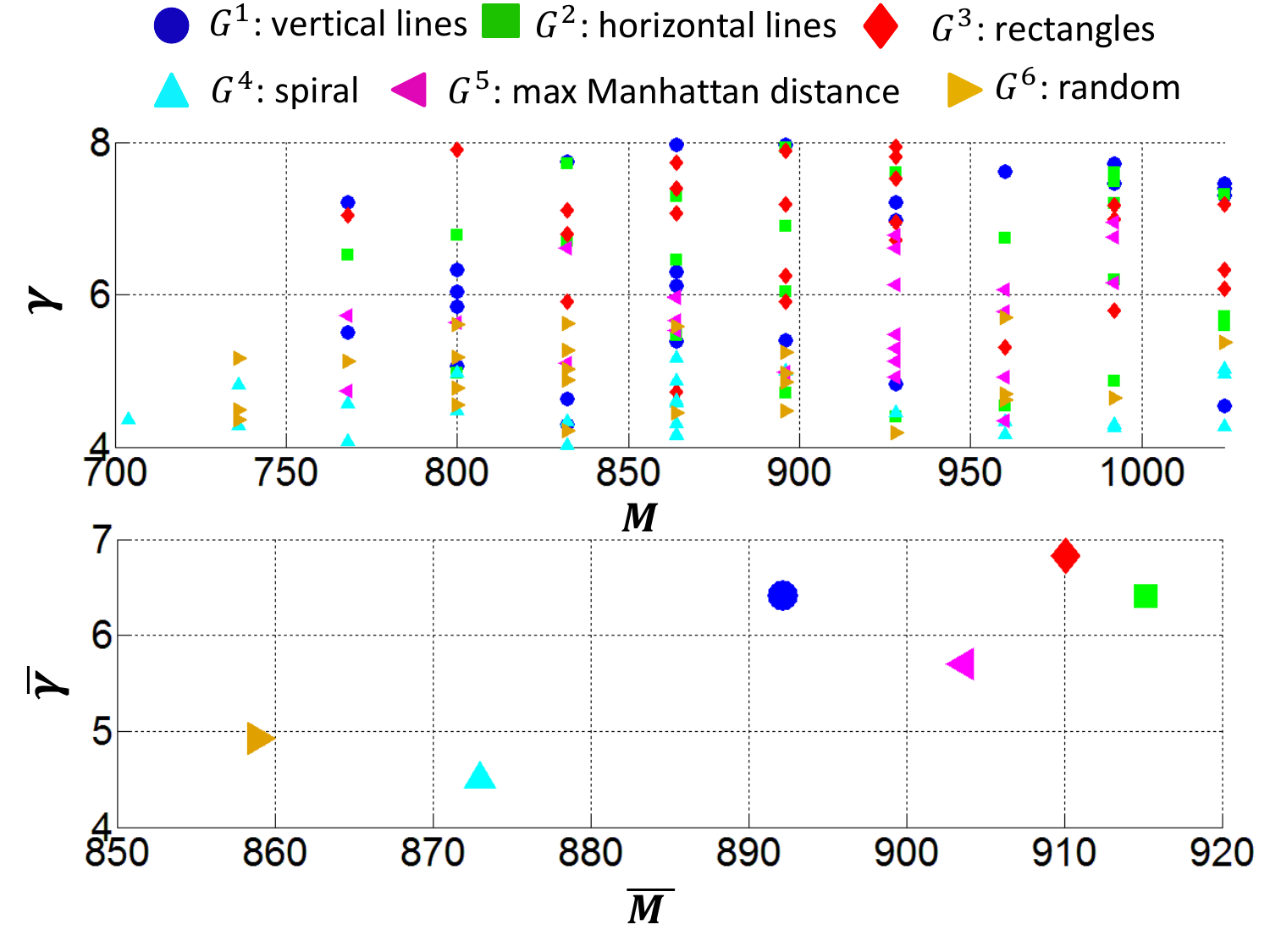}
\end{center}
\caption{Top: relationship of $M$ vs.\ $\gamma$ for the six considered group structures, for 25 small-scale ($32\times 32$ pixels) compressed images from a $160\times 160$-pixel chest MRI image (cf.~Figure~\ref{chest_MRI}); bottom: average value of $\gamma$ and $M$, averaged over the 25 segments.}
\label{figure_MRI}
\end{figure}

Figure~\ref{figure_MRI} shows the relationship between the penalty factor $\gamma(A,T,G)$  and the number $M$ of samples required for successful recovery for each of the six aforementioned group structures. Each point of the top scatter plot corresponds to a single $32\times 32$-pixel tile, while each point of the bottom scatter plot shows the average values of $\gamma$ and $M$, over all of the tiles, for each of the grouped measurement configuration. In these experiments, recovery success is defined by a normalized recovery error $NRE = \| s- \hat{s}\| /\|s\| < 0.1$  for $19$ out of $20$ draws of the measurement groups, uniformly at random. The values of $M$ tested are once again multiples of $4\cdot g=32$. The figure shows that while the group structures $G^1, G^2, G^3$ and $G^5$ demonstrate similar performance and values of $\gamma$, the group structure $G^4$ and the randomized group structure $G^5$ exhibit smaller values of $\gamma$ and lead to lower requirements on the number of measurements, which suggest the utility of $\gamma$ as a performance indicator for the Fourier domain grouped sampling schemes.

\begin{figure}[t]
\begin{center}
\includegraphics[width=5.5in] {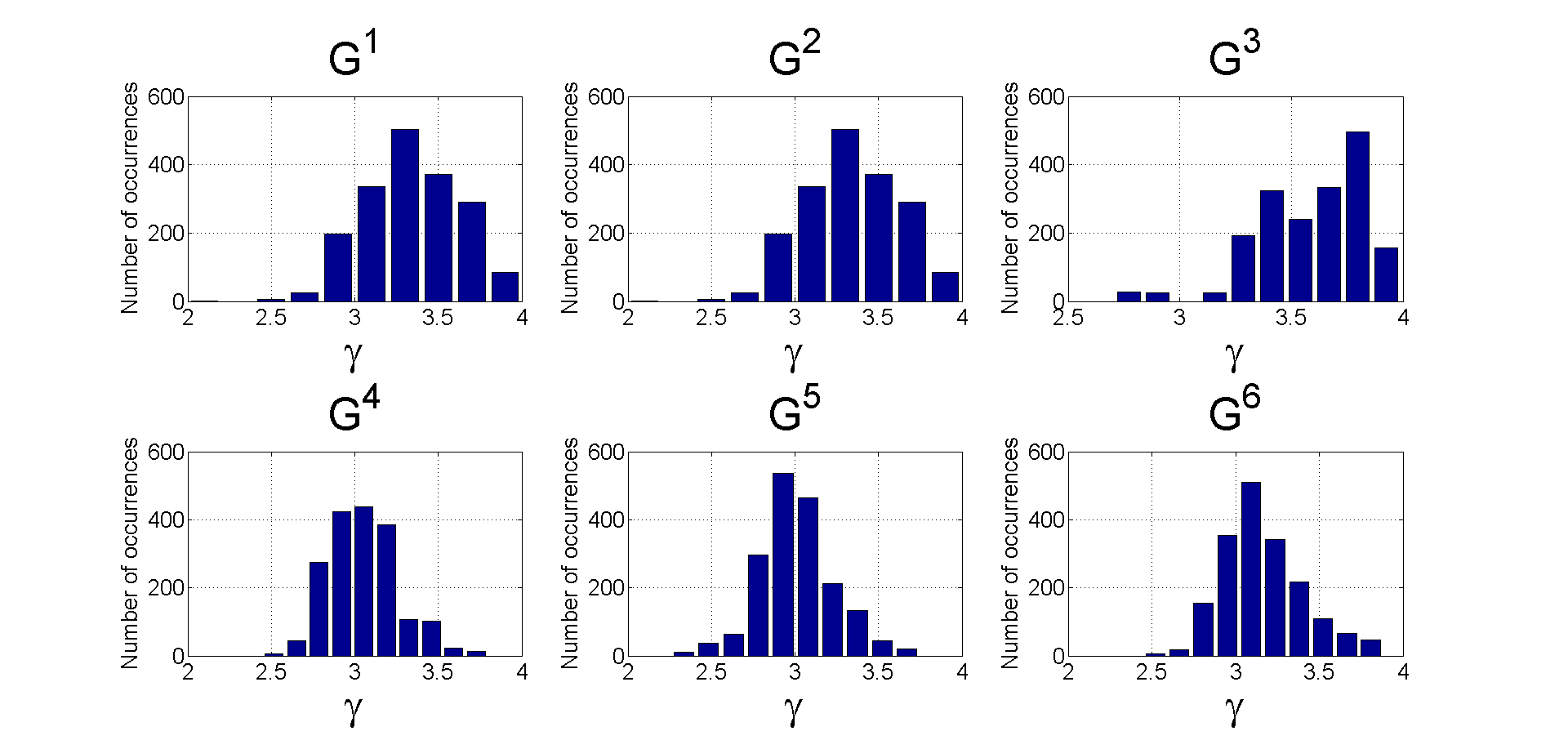}
\end{center}
\caption{Histograms of the penalty factor $\gamma$ for the transformation matrix $A$ chosen as a 2D wavelet transformation matrix of size $N^2\times N^2$, with $N=4$, for the six group structures $G$ visualized in Figure \ref{2D_groups}, for group size $g=4$, for all possible supports of size $|T|=4$.} \label{wv}
\end{figure}

\begin{figure}[t]
\begin{center}
\includegraphics[width=5.5in] {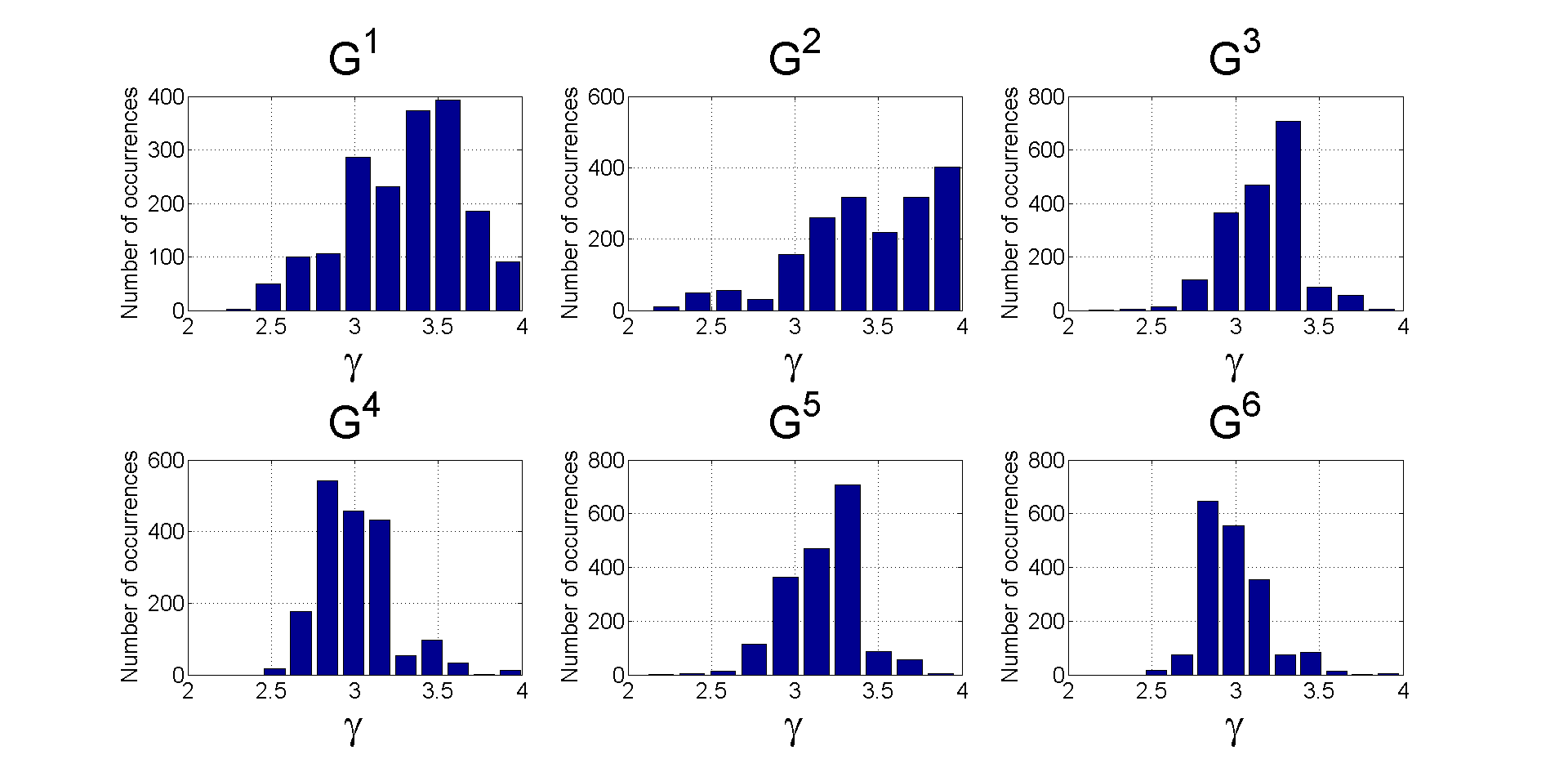}
\end{center}
\caption{Histograms of the penalty factor $\gamma$ for the transformation matrix $A$ chosen as a product of 2D Fourier and wavelet transformation matrices of size $N^2\times N^2$, with $N=4$, for the six group structures $G$ visualized in Figures \ref{2D_groups} and \ref{new_G4}, for group size $g=4$, for all possible supports of size $|T|=4$.} \label{F_wv}
\end{figure}

Figures \ref{figure_satelite} and \ref{figure_MRI} clearly indicate that the value of $\gamma$ depends on the signal support $T$.  To provide further evidence of this dependence, we present in Figures \ref{wv} and \ref{F_wv} a set of numerical results showing the ranges of values of $\gamma$ yielded by all possible choices of the signal support $T$, observed for the scenarios studied in Figures \ref{figure_satelite} and \ref{figure_MRI}. Figures \ref{wv} and \ref{F_wv} show that the values of  $\gamma$ are indeed between $\sqrt{g}$ and $g$, and that different group partitionings $G$ achieve these two extremes. Figures \ref{wv} and \ref{F_wv} also clearly indicate that the distribution of $\gamma$ depends both on $A$ and on $G$.

\section{Conclusions and Future Work}\label{conclusions}

In this work, we have presented an analytically derived multiplicative penalty on the number of measurements needed for CS recovery when the measurements exhibit grouped structure instead of the usual independently drawn measurement assumption taken by most existing CS literature. Such grouped sampling is of large practical interest as full randomization of measurements is difficult to achieve in many compressive sensing acquisition systems. We showed the utility of the introduced penalty factor as an indicator of the performance for acquisition scenarios of practical interest. A notable limitation of the introduced penalty factor $\gamma$ is that it is dependent on the signal support. We expect further work to focus on penalty metrics that are independent of the support of the signal being measured, and to expand the guarantees provided to more applicable approximately sparse signals and to noisy measurement schemes.

\section{Acknowledgements}

We thank Robert Jackson for suggesting this architecture for addressing the interference problem in wideband receivers, and Waheed Bajwa, Mark Rudelson, and Joel Tropp for helpful comments.


\appendices
\section{Proof of Theorem \ref{theorem1_2}}\label{first_proof}
Denote
\begin{equation}
Y:=\frac{N}{M}A_{\Omega T}^HA_{\Omega T}-I =\frac{N}{M}\sum_{i=1}^{N/g}\delta_i A_{G_iT}^HA_{G_iT}-I,
\end{equation}
where $\delta_i$ is a Bernoulli random variable with 
$P(\delta_i=1)=\frac{M}{N}$.
Because $A^HA=I$, we have
\begin{equation}
\sum_{i=1}^{N/g} A_{G_iT}^HA_{G_iT}=I,
\end{equation}
and so we can write
\begin{equation}
Y=\frac{N}{M} \sum_{i=1}^{N/g} \left( \delta_i -\frac{M}{N} \right)  A_{G_iT}^HA_{G_iT}  =:\sum_{i=1}^{N/g} Y_i.
\end{equation}
We will now use \cite[Theorem 1.4]{Trop11}, which we include below for completeness.
\begin{thm}\label{theorem1_4_Tropp}
Consider a finite sequence $\{Y_i\}$ of independent self-adjoint random matrices with dimension $d$. Assume that each matrix $Y_i$ satisfies $\mathbb{E}\{Y_i\}=0$  and $\left\|Y_i\right\|\leq B $ almost surely.
Then, for all $t\geq0$,
\begin{displaymath}
P\left\{ \left\|\sum_i Y_i\right\| > t \right\}\leq d\cdot \exp \left(\frac{-t^2/2}{\sigma^2+Bt/3} \right),
\end{displaymath}
\begin{displaymath}
where \quad \sigma^2=\left\|\sum_i \mathbb{E} Y_i^2\right\|.
\end{displaymath}
\end{thm}
\noindent
For our case,
\begin{equation}
Y_i=\left( \delta_i -\frac{M}{N} \right)  A_{G_iT}^HA_{G_iT} \frac{N}{M}
\end{equation}
and $E(Y_i)=0$. We find a bound $B$ on $\left\|Y_i\right\|$:

\begin{equation}\label{B_1}
\begin{split}
\left\|Y_i\right\|&=\sup_{f_1,f_2}|\langle f_1,Y_i f_2\rangle |\leq   \sup_{f_1,f_2} \left| \left\langle f_1,\frac{N}{M} \sum_{l\in G_i}a^l\otimes a^l f_2 \right\rangle \right| \leq \frac{N}{M} \sup_{f_1,f_2} \sum_{l\in G_i}|\langle f_1,a^l\otimes a^l f_2\rangle | \\
&= \frac{N}{M} \sup_{f_1,f_2} \sum_{l\in G_i}|\langle f_1,a^l\rangle \langle a^l, f_2\rangle |  \leq \frac{N}{M} \left\|a^l\right\|^2\sup_{f_2}\sum_{l\in G_i}\frac{|\langle a^l,f_2\rangle |}{||a^l||}\leq \frac{N}{M} \mu^2(A) |T| \gamma =:B,
\end{split}
\end{equation}
where the supremum is over unit-norm vectors $f_1$ and $f_2$,  $a^l$ is the $l^{th}$ row of the matrix $A_T$, and $\gamma$ is defined in \reqn{gamma}.

Next, we calculate $\sigma^2$ from Theorem \ref{theorem1_4_Tropp} as
\begin{equation}\label{sigma_thm_1}
\sigma^2=\left\|\sum_{i=1}^{N/g}\mathbb{E}(Y^2_i)\right\|= \textrm{var} \{\delta_i\}\frac{N^2}{M^2}\left\|\sum_{i=1}^{N/g}(A_{G_iT}^HA_{G_iT})^2\right\|.
\end{equation}
Since $A_{G_iT}^HA_{G_iT}$ is a Hermitian matrix, its eigendecomposition is $A_{G_iT}^HA_{G_iT}= \Omega_i \Lambda_i \Omega_i^H$, where $\Omega_i$ is a matrix whose columns are the orthonormal eigenvectors $\omega_{ij}$, $j=1,\ldots, |T|$, of the matrix $A_{G_iT}^HA_{G_iT}$, and $\Lambda_i$ is a diagonal matrix containing the eigenvalues $\{\lambda_{ij}\}_{j=1}^{|T|}$ of the matrix $A_{G_iT}A_{G_iT}^H$.  Thus, we can write
\begin{align*}
A_{G_iT}^HA_{G_iT}A_{G_iT}^HA_{G_iT} &=\Omega_i \Lambda_i \Omega_i^H\Omega_i \Lambda_i \Omega_i^H =\Omega_i \Lambda_i^2 \Omega_i^H = \sum_{j=1,\ldots, |T|} \lambda_{ij}^2 \omega_{ij} \omega_{ij}^H,
\end{align*}
and so
\begin{equation}\label{g}
\left\|\sum_{i=1}^{N/g}A_{G_iT}^HA_{G_iT}A_{G_iT}^HA_{G_iT}\right\|= \left\|\sum_{i=1}^{N/g} \sum_{j=1}^{|T|} \lambda_{ij}^2 \omega_{ij} \omega_{ij}^H\right\|.
\end{equation}
The right side of \reqn{g} is a weighted double sum of positive semidefinite matrices. The spectral norm of such a sum increases monotonically with the value of each of the weighting coefficients. Therefore, we can upper-bound \reqn{g} by replacing $\lambda_{ij}$ with $\max_{i=1,...,\frac{N}{g}} \max_ {j=1,\ldots, |T|} \lambda_{ij}$ and taking it out of the operator norm:
\begin{equation}\label{eigendecomposition}
\begin{split}
& \left\|\sum_{i=1}^{N/g}A_{G_iT}^HA_{G_iT}A_{G_iT}^HA_{G_iT}\right\|  \leq \max_{i=1,...,\frac{N}{g}} \max_{j=1,\ldots, |T|} \lambda_{ij}\left\|\sum_{i=1}^{N/g} \sum_{j=1}^{|T|} \lambda_{ij} \omega_{ij} \omega_{ij}^H\right\| \\
&\quad \quad \quad \quad=\max_{i=1,...,\frac{N}{g}} \left\|A_{G_iT}^HA_{G_iT}\right\|\cdot \left\|\sum_{i=1}^{N/g} A_{G_iT}^HA_{G_iT}\right\|.
\end{split}
\end{equation}
With \reqn{eigendecomposition} we can bound \reqn{sigma_thm_1} by
\begin{align}
\sigma^2 &\leq  \textrm{var} \{\delta_i\}\frac{N^2}{M^2} \max_{i=1,...,\frac{N}{g}} ||A_{G_iT}^HA_{G_iT}||\cdot \left\|\sum_{i=1}^{N/g} A_{G_iT}^HA_{G_iT}\right\| \nonumber\\
&= \textrm{var} \{\delta_i\}\frac{N^2}{M^2} \max_{i=1,...,\frac{N}{g}} ||A_{G_iT}^HA_{G_iT}||\cdot || A_{T}^HA_{T}||
\end{align}
With the assumption of the orthogonality of matrix $A$ ($A^HA=I$) of Theorems \ref{theorem_Candes} and  \ref{theorem_1}, we have $|| A_{T}^HA_{T}||=1$. Using the definition of the spectral norm \reqn{spec_norm_def}, we can further write
\begin{align}
\sigma^2 & \leq \frac{M}{N}\left(1-\frac{M}{N}\right)\frac{N^2}{M^2} \max_{i=1,...,\frac{N}{g}} \sup_{\left\|f_1\right\|= \left\|f_2\right\|=1} \left|\left\langle f_1, \sum_{l\in G_i} a^l \otimes a^l f_2\right\rangle \right| \nonumber\\
&\leq \frac{N}{M} \max_{i=1,...,\frac{N}{g}} \sup_{\left\|f_1\right\|= \left\|f_2\right\|=1} \sum_{l\in G_i} |\langle f_1,  a^l \otimes a^l f_2\rangle |  \nonumber\\
&= \frac{N}{M} \max_{i=1,...,\frac{N}{g}} \sup_{\left\|f_1\right\|= \left\|f_2\right\|=1} \sum_{l\in G_i} |\langle f_1,  a^l\rangle | \cdot |\langle a^l, f_2\rangle | \nonumber\\
&\leq \frac{N}{M} \left\|a^l\right\|^2 \max_{i=1,...,\frac{N}{g}} \sup_{\left\|f_1\right\|=1} \sum_{l\in G_i} \frac{\langle f_1,  a^l\rangle }{||a^l||}\nonumber\\
&\leq \frac{N}{M} \mu^2(A) |T| \max_{i=1,...,\frac{N}{g}} \sup_{\left\|f_1\right\|=1} \sum_{l\in G_i} \frac{\langle f_1,  a^l\rangle }{||a^l||}  \nonumber\\
&=\frac{N}{M} \mu^2(A) |T| \gamma =B. \label{gamma_1}
\end{align}
We put together \reqn{B_1}, \reqn{gamma_1} and Theorem \ref{theorem1_4_Tropp} to write
\begin{equation}
P(||Y||\geq1/2)\leq |T| \cdot \exp\left\{ \frac{-1/8}{7/6\cdot  \frac{N}{M} \mu^2(A)|T| \gamma }\right\},
\end{equation}
which proves Theorem \ref{theorem1_2}.

\section{Proof of Lemma \ref{lemma1}}\label{proof_lemma1}
One can express $v^0$ as
\begin{equation}
v^0=\sum_{i=1}^{N/g}\delta_i \sum_{l\in G_i} A(l, t_0)a^{l}.
\end{equation}
Now due to the orthogonality of the columns of the matrix $A$,
\begin{equation}
\sum_{i=1}^{N/g}\sum_{l\in G_i} A(l,t_0)A(l,t)=0,
\end{equation}
and we can write
\begin{equation}\label{V0}
v^0=\sum_{i=1}^{N/g}(\delta_i -E(\delta_i)) \sum_{l\in G_i} A(l,t_0)a^l=\sum_{i=1}^{N/g}Y_i,
\end{equation}
with
\begin{equation}\label{V0Yi}
Y_i:=\left(\delta_i - \frac{M}{N}\right) \sum_{l\in G_i} A(l,t_0)a^{l}.
\end{equation}
We see that $E(Y_i)=0$ and we can write
\begin{equation}\label{EV02}
\begin{split}
E||v^0||^2&=E\left\|\sum_{i=1}^{N/g} Y_i\right\|^2=E\left<\sum_{i=1}^{N/g} Y_i,\sum_{i'=1}^{N/g} Y_i\right> =E\sum_{i=1}^{N/g}\left< Y_i, Y_i\right>+E\sum_{i\neq i'}^{N/g}\left< Y_i, Y_{i'}\right>=\sum_{i=1}^{N/g}E\left< Y_i, Y_i\right>.
\end{split}
\end{equation}
Each element of the sum above can be bounded by
\allowdisplaybreaks
\begin{align}
&E\left< Y_i, Y_i\right> =E\left< \left(\delta_i - \frac{M}{N}\right) \sum_{l\in G_i} A(l,t_0)a^{l}, \left(\delta_i - \frac{M}{N}\right) \sum_{l'\in G_i} A(l',t_0)a^{l'} \right> \nonumber\\
& \quad  =\textrm{var}\{\delta_i\}\left< \sum_{l\in G_i} A(l,t_0)a^{l} ,  \sum_{l'\in G_i} A(l',t_0)a^{l'} \right> \nonumber\\
& \quad  =\frac{M}{N}\left( 1-\frac{M}{N}\right) \sum_{l\in G_i} A(l,t_0)\sum_{l'\in G_i} A(l',t_0) \left< a^{l},a^{l'} \right> \nonumber\\
& \quad  \leq \frac{M}{N} \sum_{l\in G_i} |A(l,t_0)|\sum_{l'\in G_i} |A(l',t_0)| \cdot \left|\left< a^{l},a^{l'} \right>\right| \nonumber\\
& \quad  \leq \mu(A) \frac{M}{N}\sum_{l\in G_i} |A(l,t_0)|\sum_{l'\in G_i}  \left|\left< a^{l},a^{l'} \right>\right| \nonumber\\
& \quad \leq \mu(A) \frac{M}{N} \sum_{l\in G_i} |A(l,t_0)| \sum_{l' \in G_i} \frac{\left| \left\langle a^l, \frac{a^{l'}}{\left\| a^{l'} \right\|} \right\rangle \right|}{\left\| a^{l} \right\|} \cdot \left\|a^{l}\right\|\cdot \left\|a^{l'}\right\| \nonumber\\
& \quad  \leq\mu^3(A)|T|\frac{M}{N} \sum_{l\in G_i} |A(l,t_0)| \sum_{l' \in G_i} \frac{\left| \left\langle a^l, \frac{a^{l'}}{\left\| a^{l'} \right\|} \right\rangle \right|}{\left\| a^{l} \right\|} \nonumber\\
& \quad  \leq \mu^3(A)|T|\frac{M}{N} \gamma  \sum_{l\in G_i} |A(l,t_0)|. \label{EV03}
\end{align}
Putting together \reqn{EV02} and \reqn{EV03}, we get
\begin{align*}
 E||v^0||^2&\leq \frac{M}{N} \mu^3(A) |T| \gamma  \sum_{i=1}^{N/g} \sum_{l'\in G_i} |A(l',t_0)| =\frac{M}{N}\mu^3(A)|T| \gamma \sum_{l=1}^N |A(l,t_0)|=\frac{M}{N}\mu^3(A)|T|\gamma \left\| A(:,t_0) \right\|_1.
\end{align*}
Now since $A^HA=I$, we have $\left\| A(:,t_0) \right\|_2=1$, and since for any vector $h$ of length $N$ we have $||h||_1\leq \sqrt{N}||h||_2$, it follows that
\begin{equation}
E||v^0||^2\leq \frac{M}{\sqrt{N}} \mu^3(A) |T| \gamma,
\end{equation}
which proves Lemma \ref{lemma1}.

\section{Proof of Lemma \ref{lemma2}}\label{proof_lemma2}
By definition,
\begin{equation}
||v^0||=\sup _{||f||=1} \left<v^0,f\right>=\sup _{||f||=1} \sum_{i=1}^{N/g} \left<Y_i,f\right>,
\end{equation}
with $v^0$ from \reqn{V0} and $Y_i$ from \reqn{V0Yi}.
For completeness, we reproduce below \cite[Theorem 3.2]{Cand06}, which we use to prove Lemma \ref{lemma2}.
\begin{thm} \label{thm32cand} Let $Y_1, \ldots, Y_N$ be a sequence of independent random variables taking values in a Banach space and let $Z$ be the supremum $Z=\sup_{f\in \cal{F}}\sum_{n=1}^N f\left( Y_i \right)$, where $\cal{F}$ is a countable family of real-valued functions. Assume that $|f(Y)|<B$ for every $f\in \cal{F}$ and all $Y$, and $\mathbb{E} f(Y_i)=0$ for every $f\in \cal{F}$ and $i=1,\ldots,N$. Then, for all $t\geq 0 $,
\begin{equation*}
P\left(\left| Z- \mathbb{E}Z  \right| >t\right)\leq 3\exp \left( \frac{-t}{KB}\log \left(1+ \frac{Bt}{\sigma^2+B \mathbb{E}\bar{Z}}\right) \right),
\end{equation*}
where
\begin{align*}
\sigma^2 &=\sup_{f\in\cal{F}} \sum_{i=1}^N \mathbb{E} f^2(Y_i),\\
\bar{Z}&=\sup_{f\in \cal {F}} \left| \sum_{i=1}^N f(Y_i)\right|,
\end{align*}
and $K$ is a numerical constant.
\end{thm}
\noindent Denote the mapping $\left<Y_i, f\right>$ for a fixed unit vector $f$ as $f(Y_i)$, so that $\bar{Z}=\sup_{||f||=1}\sum_{i=1}^{N/g}f(Y_i)=||v^0||$. We have $\mathbb{E}\{f(Y_i)\}=0$, and
\begin{align}
|f(Y_i)|&=\left|\left<\left(\delta_i - \frac{M}{N}\right) \sum_{l\in G_i} A(l,t_0)a^{l},f\right>\right|\nonumber\\
&=\left|\left(\delta_i - \frac{M}{N}\right)\sum_{l\in G_i} A(l,t_0)\left<a^{l},f\right>\right| \nonumber\\
&<\left|\sum_{l\in G_i} A(l,t_0)\left<a^{l},f\right>\right|\nonumber\\
&\leq \max_{l \in G_i} |A(l,t_0)|  \cdot \max_{l \in G_i} ||a^l|| \cdot \sum_{l \in G_i} \frac{|\langle a^l, f \rangle|}{||a^l||} \nonumber\\
&\leq \gamma \cdot\mu^2(A)\cdot \sqrt{|T|}=:B. \label{EV4}
\end{align}
Now we find a bound on $\mathbb{E} \{f^2(Y_i)\}$:
\begin{align*}
\mathbb{E}\{f^2(Y_i)\}&=\mathbb{E} \left\{\left|\left<\left(\delta_i - \frac{M}{N}\right) \sum_{l\in G_i} A(l,t_0)a^{l},f\right>\right|^2\right\} =\textrm{var}\{\delta_i\}\cdot\left|\sum_{l\in G_i}A(l,t_0)\langle a^l, f\rangle\right|^2  \\
&=\frac{M}{N}\left(1 - \frac{M}{M}\right)\cdot\left|\sum_{l\in G_i}A(l,t_0)\left< a^l, f\right>\right|^2 \leq \frac{M}{N} \mu^2(A) \sum_{l\in G_i} \left| \langle a^l, f\rangle \right|^2,
\end{align*}
and so
\begin{displaymath}
\sum_{i=1}^{N/g} \mathbb{E}\{f^2(Y_i)\}\leq  \mu^2(A) \frac{M}{N} \sum_{i=1}^{N/g}  \sum_{l\in G_i} |\left< a^l, f\right>|^2.
\end{displaymath}
We know that $\sum_{i=1}^{N/g}\sum_{l\in G_i} |\langle a^l, f\rangle|^2=1$; therefore,
\begin{align}
\sum_{i=1}^{N/g} \mathbb{E}\{f^2(Y_i)\} \leq \frac{M}{N} \mu^2(A)=:{\sigma}^2. \label{EV5}
\end{align}
Plugging \reqn{EV4} and \reqn{EV5} in Lemma \ref{lemma1}, we have
\begin{equation}
\mathbb{E}\{\bar{Z}\}= \mathbb{E}\{||v^0||\}\leq \mu^{3/2}(A) \sqrt{|T|} \sqrt{\gamma} \frac{\sqrt{M}}{N^{1/4}}.
\end{equation}
Assume that\\
\begin{displaymath}
\frac{\sqrt{\gamma}{|T|}\mu^{3/2}(A)N^{3/4}}{\sqrt{M}}<1 \ \ ; \ \   0<a\leq \frac{\sqrt{M}}{\sqrt{\gamma} \mu(A) \sqrt{N} \sqrt{|T|}};
\end{displaymath}
then with \reqn{define_sigma}, we have
\begin{equation}
\begin{split}
\bar{\sigma}^2 &= \gamma \mu^2(A) \frac{M}{N} > B \mathbb{E} \{\bar{Z}\} = B \mathbb{E} \{\left\| v^0 \right\|\} > B \mu^{3/2}(A) \sqrt{\gamma} \sqrt{|T|} \frac{\sqrt{M}}{N^{1/4}},
\end{split}
\end{equation}
and by writing $t=a \bar{\sigma}$ we have
\begin{equation}\label{t_a_sigma}
Bt=B a \bar{\sigma}\leq \bar{\sigma}^2.
\end{equation}
For $\frac{\sqrt{\gamma}{|T|}\mu^{3/2}(A) N^{3/4}}{\sqrt{M}}>1$ and $0<a\leq \left(\frac{{M}}{{\gamma} \mu(A) \sqrt{N}}\right)^ {1/4}$, with \reqn{define_sigma} we have
\begin{align}
\bar{\sigma}^2&=\gamma^{3/2}\mu^{7/2}(A) |T|\sqrt{M}= B \mathbb{E} \{\bar{Z}\} = B \mathbb{E} \{\left\| v^0 \right\|\}\nonumber = B\mu^{3/2}(A)\sqrt{\gamma}  \sqrt{|T|} \frac{\sqrt{M}}{N^{1/4}}
\end{align}
and so
\begin{equation}\label{Bt_sigma}
Bt\leq \bar{\sigma}^2.
\end{equation}
Putting together \reqn{t_a_sigma}, \reqn{Bt_sigma}, and Theorem \ref{thm32cand}, we can write
\begin{equation*}
P\left(||v^0||> \mu^{3/2} (A) N^{-1/4} \sqrt{\gamma M |T|}+a\bar{\sigma}\right)< 3 e^{-\kappa a^2},
\end{equation*}
where $\kappa$ is a numerical constant $\kappa = \frac{\log(1.5)}{K}$ and $K$ comes from Theorem \ref{thm32cand}. This completes the proof of Lemma \ref{lemma2}.

\section{Proof of Lemma \ref{lemma3}}\label{proof_lemma3}

Denote the events
$$E_1: \  \left\{ \left\| A^H_{\Omega T} A_{\Omega T}\right\| \geq \frac{M}{2N}\right\}$$
and
$$E_2: \ \left\{  \sup_{t_0\in T^c} ||v^0||\leq \mu^{3/2} (A) N^{-1/4} \sqrt{\gamma M |T|}+a \bar{\sigma}\right\}.$$
We can write
\begin{align*}
P\left(\sup_{t_0\in T^c} ||w^0|| \geq 2  N^{3/4} \mu^{3/2}  \sqrt{\frac{\gamma |T|}{M}}+\frac{2 N a\bar{\sigma}}{M} \right) & \leq P ( \overline{ E_1 \cap E_2} )= P ( \overline{E_1} \cup \overline{E_2})\leq P (\overline{E_1})+P(\overline{E_2}),
\end{align*}
and with Lemma \ref{lemma2} we have
\begin{equation}
\begin{split}
P & \left(\sup_{t_0\in T^c}||w^0|| \geq 2  N^{3/4} \mu^{3/2}  \sqrt{\frac{\gamma |T|}{M}}+\frac{2 N a\bar{\sigma}}{M} \right) \leq  P\left( ||A_{\Omega T}^HA_{\Omega T}|| \leq \frac{M}{2N} \right)+3  e^{-\kappa a^2},
\end{split}
\end{equation}
which proves Lemma \ref{lemma3}.

\end{document}